\documentclass[aps,prl,superscriptaddress,showpacs,preprint
]{revtex4-1}

\usepackage[utf8]{inputenc}

\usepackage[english]{babel}
\usepackage[range-phrase=~to~,range-units=single]{siunitx}
\usepackage{graphicx}
\usepackage{nicefrac}
\usepackage{amsmath}
\usepackage{xspace}

\usepackage{hyperref}

\newcommand{\2}{\mbox{$2H$-TaSe$_2$}}
\newcommand{\rt}{room~temperature\xspace}

\newcommand{\figref}[1]{Fig.\,\ref{#1}\xspace}
\newcommand{\tmdc}{transition-metal di\-chalco\-genide}
\newcommand{\tmdcs}{transition-metal di\-chalco\-genides}
\newcommand{\eels}{Electron Energy-Loss Spectroscopy}

\renewcommand{\epsilon}{\varepsilon}
\newlength{\myfigwidth}
\newcommand{\revision}[1]{#1}

\begin{document}

\title{Plasmon Evolution and Charge-Density Wave Suppression in Potassium Intercalated $2H$-TaSe$_2$}
\date{\today}

\author{A.\,König}
\email[E-mail: ]{a.koenig@ifw-dresden.de}
\affiliation{IFW Dresden, Institute for Solid State Research, P.\,O. Box 270116, D-01171 Dresden, Germany}
\author{K.\,Koepernik}
\affiliation{IFW Dresden, Institute for Solid State Research, P.\,O. Box 270116, D-01171 Dresden, Germany}
\author{R.\,Schuster}
\affiliation{IFW Dresden, Institute for Solid State Research, P.\,O. Box 270116, D-01171 Dresden, Germany}
\author{R.\,Kraus}
\affiliation{IFW Dresden, Institute for Solid State Research, P.\,O. Box 270116, D-01171 Dresden, Germany}
\author{M.\,Knupfer}
\affiliation{IFW Dresden, Institute for Solid State Research, P.\,O. Box 270116, D-01171 Dresden, Germany}
\author{B.\,Büchner}
\affiliation{IFW Dresden, Institute for Solid State Research, P.\,O. Box 270116, D-01171 Dresden, Germany}
\author{H.\,Berger}
\affiliation{Institut de Physique de la Mati\`ere Condens\'ee, EPFL, CH-1015 Lausanne, Switzerland}


\begin{abstract}
We have investigated the influence of potassium intercalation on the formation of the charge-density wave (CDW) instability in \2 by means of \eels\ and density functional theory. Our observations are consistent with a filling of the conduction band as indicated by a substantial decrease of the plasma frequency in experiment and theory. In addition, elastic scattering clearly points to a destruction of the CDW upon intercalation as can be seen by a vanishing of the corresponding superstructures. This is accompanied by a new superstructure, which can be attributed to the intercalated potassium. Based on the behavior of the $c$-axis upon intercalation we argue in favor of interlayer-sites for the alkali-metal and that the lattice remains in the $2H$ modification.
\end{abstract}

\maketitle

\section{Introduction}
\label{sec:intro}

Due to their layered structure, \tmdcs\ (TMDCs) show a variety of anisotropic properties, such as thermal expansion, sound velocity and thermal as well as electrical conductivity~\cite{Brown1965, Quinn1966, vanMaaren1967, Wilson1969}. The lowered dimensionality often also manifests itself in the occurrence of a charge-density wave (CDW) phase transition within the $ab$-plane of the crystal. In the present example of \2, this CDW superstructure is incommensurate below \SI{120}{K} and becomes commensurate going below \SI{90}{K}~\cite{Wilson1975, Moncton1975, Moncton1977}. A detailed study of the phase transition---containing an incommensurate superstructure followed by a lock-in transition to a commensurate ordering vector---has for example been realised by Neutron scattering~\cite{Moncton1977} or x-ray diffraction~\cite{Leininger2011}. This transition as well as the CDW state itself were also investigated by angle-resolved photoemission measurements~\cite{Borisenko2008, Inosov2009}, clearly showing the opening of a gap at the Fermi level when going to the ordered phase. However, the reasoning that the Fermi surface is partially gapped already for temperatures far above the phase transition and the link of this fact to a possible nesting mechanism for the CDW order is supported by reflectivity as well as resistivity measurements~\cite{Vescoli1998}.

Recent spectroscopic investigations on \2 showed results seemingly contradicting the conventional description of a metal---apart from the observation of the partial gapping---namely, a negative slope of the plasmon energy dispersion, which can be assigned to an interplay of the charge density fluctuations and the plasma resonance~\cite{vanWezel2011a}.

While the $ab$-planes of the crystals of the TMDCs consist of hexagonal layers, the $c$-direction is formed by sandwich-like structures of these layers, bound to each other through \emph{van-der-Waals-}interaction. The gaps between the building blocks are the typical place for all kinds of intercalates to arrange within the crystal~\cite{Friend1987,Rouxel1979}.
Thereby, the range of possible doping substances is as wide as the spectrum of physical properties to tune with intercalation.

From the electronic point of view it is reported that the band structure changes extensively upon alkali metal (Na with Na/Ta$ = 0.3$ and Cs with Cs/Ta varying between 0.3 and~0.6) doping in \2~\cite{Brauer2001, Pettenkofer1992} and related compounds, such as $1T$-Rb$_x$TaSe$_2$~\cite{Stoltz2003} and $1T$-Cs$_x$TaSe$_2$ ( up to $x=0.5$)~\cite{Crawack2000}.

Structural changes upon intercalation can be observed in va\-rious poly\-types of the TMDCs. Primarily the $c$-axis is widened, while the
spacing within the $ab$-plane remains nearly unchanged~\cite{Starnberg2000, Omloo1970}. Slight changes in the in-plane crystal structure
indicated by a change of super structure are observed for different compounds such as the $1T$-Cs$_x$TaSe$_2$ mentioned
before~\cite{Crawack2000} as well as in K$_x$TiS$_2$~\cite{Pronin2001}. Because of the suppression of the CDW, as for example in
$2H$-K$_x$TaS$_2$~\cite{Biberacher1985}, the observed super structures can be assumed to result from the intercalated atoms.

Structural changes are also discussed in terms of staging effects in the class of silver doped TMDCs in close analogy to intercalated graphite
compounds. The added silver atoms are supposed to slightly disarrange the chalcogenide atoms and with it opening new scattering channels. The
observed stages are at doping rates of one and two thirds, respectively, for Ag$_x$TaS$_2$ as well as Ag$_x$TiS$_2$~\cite{Scholz1980}. Other
examples also show super structure changes in Ag$_x$NbS$_2$~\cite{Wiegers1988,vanderLee1991}.

We report on the \emph{in situ} intercalation into the bulk sample of \2 with the alkali metal potassium, reaching a comparatively high stable doping rate. \eels\ is applied to investigate changes in the electronic as well as crystal structure.
We demonstrate that doping levels up to K$_{0.77}$TaSe$_2$ can be reached which are characterized by well defined plasmon excitations, while the CDW superstructure is suppressed upon doping.

\section{Experiment and Calculations}
\label{sec:experimental}

\begin{figure}[ht]
  \centering
  \includegraphics[width=.92\textwidth]{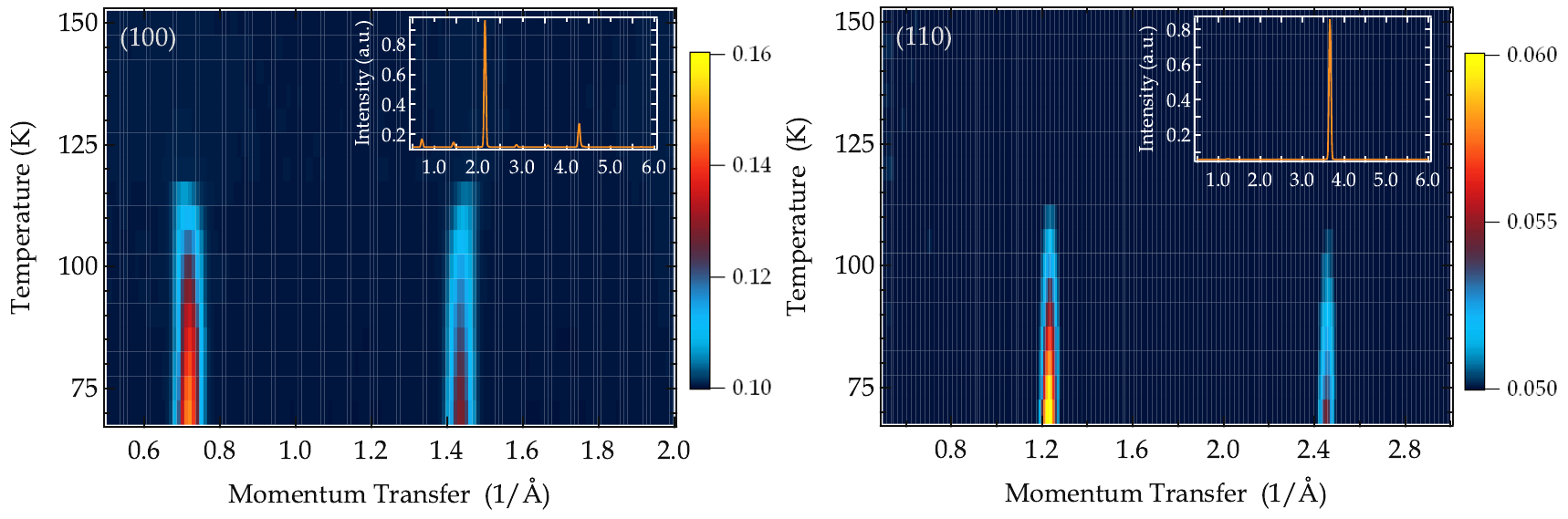}
  \caption{Charge density wave superstructure reflections arising by cooling down below the phase transition at \SI{120}{K}. The figures show a part of the electron diffraction profile, while the corresponding insets show a bigger range of momentum transfer with the prominent Bragg reflections \revision{as well as the superstructures} of the basic crystallographic structure of \2 \revision{at low temperatures}.}
\label{fig:superstruct}
\end{figure}

Single crystals of \2 were grown from Ta metal and Se (\SI{99.95}{\percent} and \SI{99.999}{\percent} purity, respectively) by iodine vapor transport in a gradient of \SIrange{600}{620}{\degreeCelsius}, the crystals growing in the cooler end of the sealed quartz tubes. A very slight excess of Se was included (typically \SI{0.2}{\percent} of the charge) to ensure stoichiometry in the resulting crystals. Each experimental run lasted for \SIrange{250}{300}{h}. This procedure yielded single crystals with maximum size of $10 \times 10 \times \SI{0.2}{\cubic\milli\meter}$. Appropriate thin films (thickness of about \SI{100}{nm}) were prepared either by the use of an ultramicrotome or by cleaving with adhesive tape, whereby the crystal was repeatedly cleaved and the glue layer dissolved afterwards. The films were mounted onto standard electron microscopy grids and transferred to the spectrometer. As depicted in \figref{fig:superstruct}, a very good sample quality is achieved, since discrete Bragg reflections as well as clear superstructures of the CDW arising at the predicted transition temperature (\SI{120}{K}) are visible.

Potassium intercalation was achieved by evaporation from commercial SAES (SAES GETTERS S.\,P.\,A., Italy) getter sources under ultra high vacuum conditions (base pressure below \SI{e-10}{mbar}). The intercalation levels were achieved by several discrete intercalation steps, of which each took about \SI{3}{min}. The relation between intercalation time and corresponding increase in potassium content could not be shown to be linear due to several reasons, such as varying getter quality and inhomogeneous saturation of the sample. Furthermore, the saturation potassium content was found to be stable over a time span of several months at \rt\ and ultra-high vacuum conditions.
All measurements were carried out with an \SI{172}{kV} electron energy-loss spectrometer equipped with a He flow cryostat. The resolution of the spectrometer is \SI{0.03}{\per\angstrom} and \SI{80}{meV} for momentum and energy-loss, respectively. The spectrometer setup can be found in Ref.\,\cite{Fink1989}. For the present studies, the access to elastic scattering processes by measuring at zero energy-loss is of great importance.

\begin{figure}[ht]
  \centering
  \includegraphics[width=0.8\textwidth]{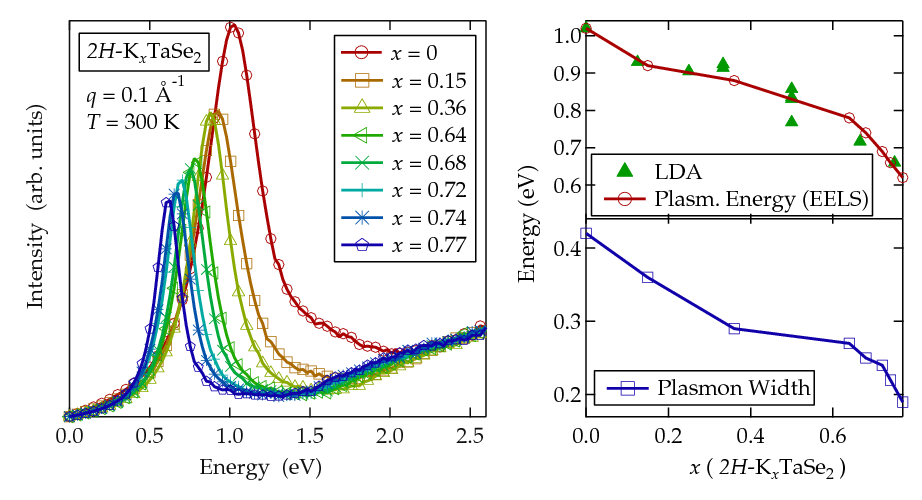}
  \caption{Loss-function at constant momentum transfer \revision{$q$} showing a red~shift of the \SI{1}{eV} plasma resonance for increasing potassium content (left panel). The right panel depicts the plasmon energy as well as the width of the plasmon peak, corresponding to the left panel, with respect to potassium content $x$ per TaSe$_2$ unit as estimated with the help of LDA calculations (cf. Fig.\,\ref{fig:Theoretical-results:-a:}c).}
\label{fig:pot_loss}
\end{figure}

To gauge and analyse the experimental results we performed density functional theory (DFT) calculations within the local-density-approximation (LDA). We used the scalar relativistic mode of the full potential local orbital code (FPLO)\cite{FPLO99} in version 9.09. In order to model the intercalation we used two approaches. In the virtual crystal approximation (VCA) K$_{18+x}$-layers are inserted between adjacent TaSe$_{2}$ layers, where $18+x$ denotes the number of protons and electrons on the VCA-pottasium atom, which models the VCA equivalent of K$_{x}$TaSe$_{2}$. In this way the filling effect of the TaSe bands upon doping can be described. For judging the quality of VCA and for incorporating more realistic Fermi surface reconstruction effects we constructed several supercells (SC) with a corresponding K/Ta ratio. \revision{Guided by the experimentally observed superstructure patterns we focus on simple supercells with at least a threefold axis. For $x=\nicefrac{1}{2}$ we also considered an orthorhombic cell.} Especially for small concentrations it is not obvious that pottasium intercalates evenly into all interlayers. To cover such effects two series of supercells were created, one with potassium distributed as evenly as possible and one, where only every second interlayer contains the dopant \revision{(for $x=\nicefrac{1}{8},\nicefrac{1}{3},\nicefrac{1}{2}$). Details about the superstructures are found in the supplementary material. We did not try to determine the true structural ground state, since this could become rather involved and the purpose of the supercells is merely to supplement and check the VCA calculations.} For all VCA and SC calculations we relaxed all lattice and Wyckoff parameters to accomodate breething of the lattice due to K-intercalation. The $\boldsymbol{k}$-integration for the self consistent calculations was performed via the linear tetrahedron method with a $12\times12\times4$ mesh for the simple VCA K$_{2\left(1-x\right)}$Ta$_{2}$Se$_{4}$ unit cell. For the supercell calculations the number of mesh points was adjusted accordingly to obtain an equivalent mesh density. Optical properties were calculated with a $32\times32\times24$ mesh within the whole Brillouin zone, which was checked for convergence.

\revision{For the undoped compound we performed full-relativistic calculations to check the influence of spin orbit coupling on the plasmon frequencies. It turns out that the spin-orbit-splitting of the bands at the Fermi level is quite sizable (\SI{400}{meV}) but the splittings (which reduce the $z$-dispersion) occur in such a way that their effects on the $xy$-plane plasmon energies average out over the Brillouin zone. The resulting changes of in-plane $\omega_{p}$ are negligible, while the out-of-plane plasmon energy gets reduced. We only focus on the in-plane plasmon in this work and hence used scalar relativistic calculations throughout this work.}

\section{Results and Discussion}
\label{sec:results}

\begin{figure}
\center
\includegraphics[width=\myfigwidth]{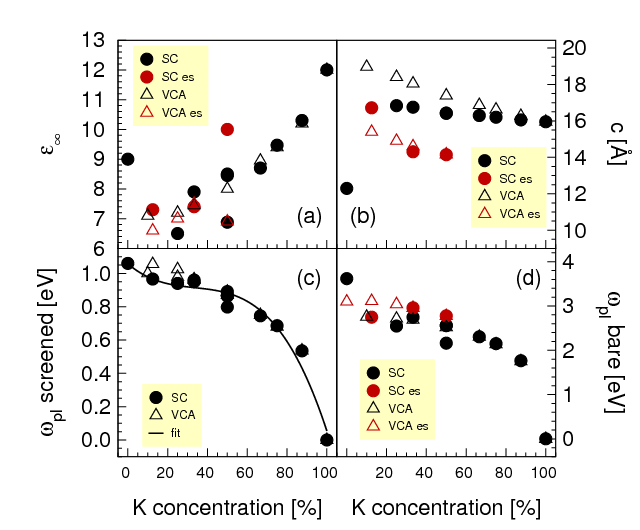}
\caption{Theoretical results of the LDA calculations (virtual crystal approximation (VCA) as well as constructed supercells (SC)): a) $\varepsilon_{\infty}$ from the interband contributions; b) the $c$-axis lattice parameter; c) the plasmon peak position of the calculated loss function, the solid line shows a fit of the supercell frequencies, used for fixing the potassium content of the measured sample; d) the bare plasmon frequency. \emph{es} denotes cells with alternatingly empty interlayers.}
\label{fig:Theoretical-results:-a:}
\end{figure}

Figure\,\ref{fig:pot_loss} (left panel) shows loss spectra of $2H$-K$_x$TaSe$_2$ for increasing potassium content~$x$. All spectra are normalized in the region around \SI{2.5}{eV} and are furthermore corrected by fitting and subtracting the elastic line, contributing at around zero energy-loss~\cite{Schuster2009}. The undoped sample shows a dominating peak at around \SI{1}{eV} that, with increasing  potassium content, shifts to lower energies. Furthermore, the normalized peak maximum as well as the peak width are reduced. \revision{This spectral feature represents the charge carrier plasmon of the investigated samples, and therefore contains information on the charge carrier density. The reduction of the plasmon energy is a direct consequence of the charge carrier density variation induced by the potassium doping. Undoped \2 is characterized by half filled conduction bands, and addition of either electrons or holes will lower the plasmon energy as has been discussed previously based on rigid band considerations \cite{Campagnoli1979}. In general, the plasmon energy therefore can be also used to analyze the doping level. Since the band structure of \2 is rather complex and variations beyond a rigid band approach cannot be excluded, we compared calculated screened plasmon energies (Fig.\,\ref{fig:Theoretical-results:-a:}, below) to the measured ones and deduced the $x$-axis in the right panel of Fig.\,\ref{fig:pot_loss}. We used the thus defined concentrations in the rest of the paper. Note, however, that the calculated and measured screened plasmon energies are remarkably similar, which makes such a gauging procedure feasible (see also discussion below).}

It can be concluded that  saturation of potassium doping is likely to be reached at about $2H$-K$_{0.77}$TaSe$_2$. Older reports on intercalation of \tmdcs\ were able to find also higher dopant rates up to $x=1$.~\cite{Friend1987,Rouxel1979} However, Rouxel~\cite{Rouxel1979} also emphasized that a homogeneous structure in the selenides could only be obtained for rates of $x=0.6$ to $0.7$. \revision{Furthermore, the spectral width of charge carrier plasmons in most cases is predominantly determined by damping via interband transitions~\cite{Paasch1970,Sturm1976}. As a consequence, we attribute the reduction of the plasmon width upon doping mainly to its shift to lower energies, where less interband transitions are available for a damping of the plasmon excitation. In addition, mid-infrared transitions as observed for undoped \2 are suppressed upon doping~\cite{Vescoli1999}, which could also reduce the plasmon damping.}

\begin{figure}[ht]
  \centering
  \includegraphics[width=\myfigwidth]{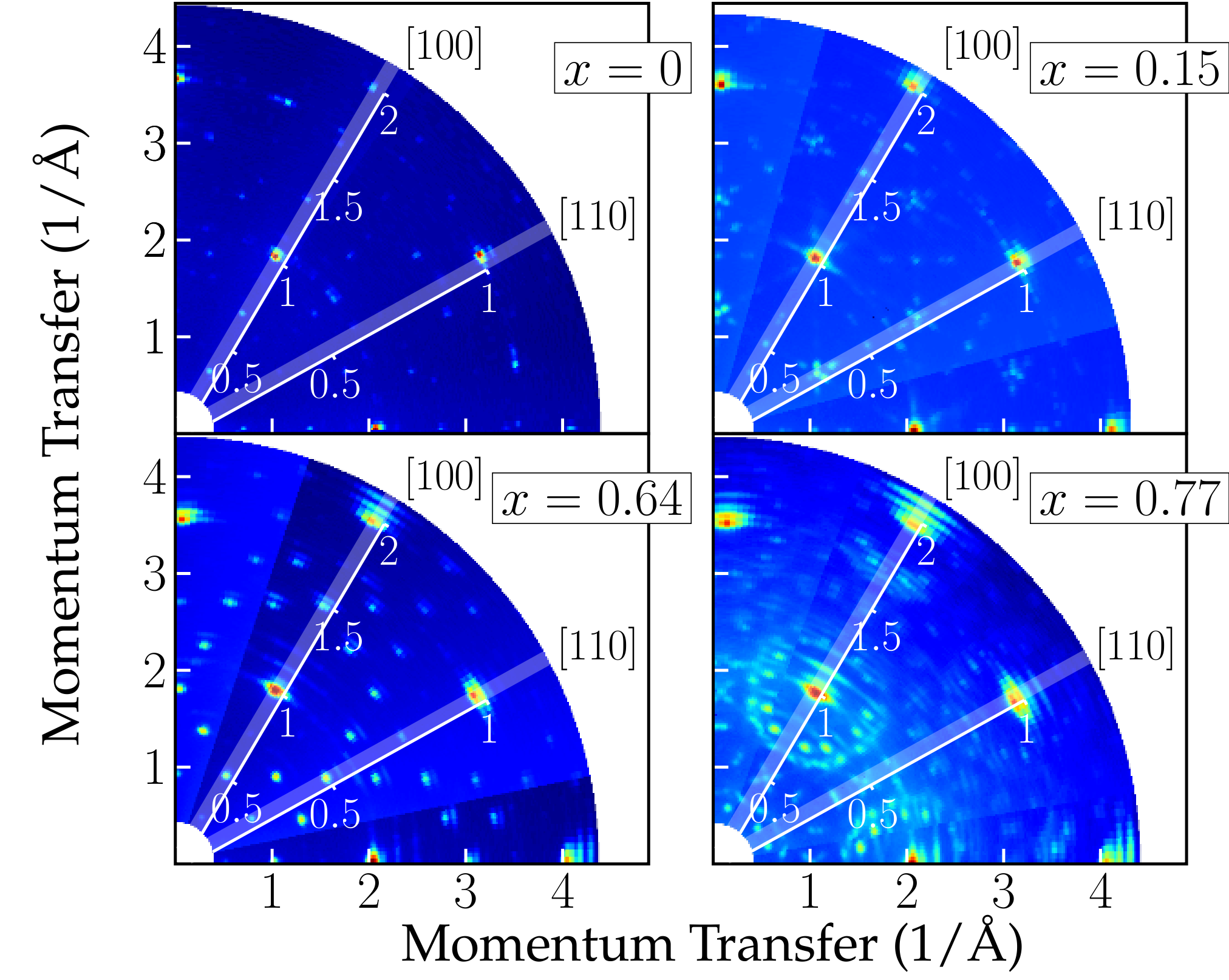}
  \caption{Elastic scattering measurements of the $ab$-plane with increasing potassium content from top left to bottom right showing CDW superstructures for low temperatures and low K contents and arising dopant superstructures for increasing K contents.}
  \label{fig:struct_doped_30020}
\end{figure}

In \figref{fig:struct_doped_30020} intensity maps, achieved by elastic scattering in EELS, are shown. They show congruent sections of the $ab$-plane with increasing potassium content from top ($x=0$) to bottom ($x=0.77$) at a temperature of \SI{20}{K}. The bright spots in the highlighted $\Gamma M$ and $\Gamma K$ direction ([100] and [110], respectively) represent the basic structure of the compound. Additional reflections at one-third and two-third of the [100] vector are observable in the undoped phase (topmost left part of \figref{fig:struct_doped_30020}), which can be assigned to the occurring CDW at temperatures below \SI{120}{K} (see also \figref{fig:superstruct}).~\cite{Wilson1975, Moncton1975, Moncton1977}

\begin{figure}[ht]
  \centering
  \includegraphics[width=\myfigwidth]{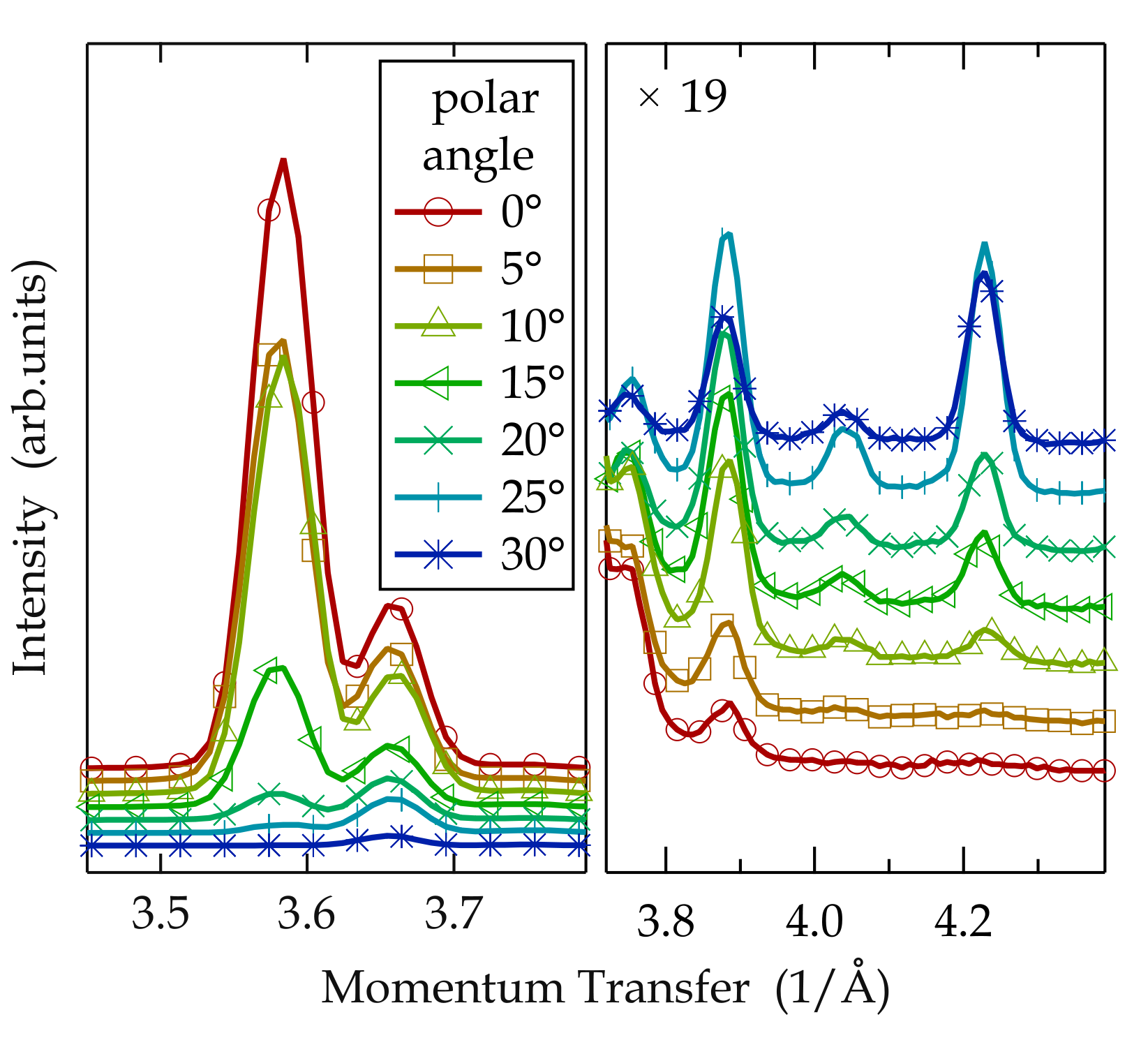}
  \caption{Elastic scattering measurements of the fully doped sample ($2H$-K$_{0.77}$TaSe$_2$) around the [110] zero reflection (left panel) showing the suppressed intensity for increasing angles of the scattering direction with respect to the $ab$-plane. The right panel shows the according [11$l$] reflections (see text), that emerge with increasing angles, allowing a calculation of the $c$-axis length, which is widened compared to the undoped crystal structure.}
  \label{fig:polar_angle}
\end{figure}

During the successive doping process several changes in the scattering image can be observed. The most prominent superstructure is a hexagon like ring at around \SI{50}{\percent} doping (bottom left part of \figref{fig:struct_doped_30020}). Upon increasing potassium addition, in this state the sample first shows a mixed superstructure of the CDW spots at one-third and two-third as well as new superstructures occuring at one-half (for $x=0.15$) and one and three quarter (for $x=0.64$), respectively. However, the latter doping state is no longer showing the original triple CDW superstructure. At the highest possible doping level (for $x=0.77$) again a ring like structure around the [100] reflection can be found.

These findings lead to the conclusion that in \2 the CDW is suppressed by K intercalation, as it was also found for the related compound $2H$-TaS$_2$~\cite{Shen2007,Hu2007,Fang2005}. At the same time new features arise that must be assigned to an increasing potassium filling of the crystal structure. At this point a comparison between alkali metal intercalated graphite and TMDCs shows striking similarities. Both material classes show a van-der-Waals coupled hexagonal layer structure and metallic behavior in their \rt\ phase~\cite{Dresselhaus2002}. Furthermore, a transition to a superconducting phase could be induced or supported, respectively, by means of alkali metal doping~\cite{Hannay1965,Fang2005}. Upon intercalation, most commonly Rb or K, graphite is able to form distinct phases, which can be distinguished as stage 1 and 2 (C$_8$X and C$_{14}$X, respectively, with X being the alkaline metal). With increasing content of the dopant it is showing a diversity of superstructures that can be assigned to the intercalant~\cite{Ruedorff1954,Nixon1968,Kambe1980,Zabel1980}. However, the formation of distinct phases present in intercalated graphite could not be observed in \2. In particular the continuous variations in line shape and position of the charge carrier plasmon do not signal the simultaneous presence of two or more phases at any doping level. \revision{Nevertheless, the appearance of new superstructure peaks at one-half of the basic unit length of \2 would be consistent with the potassium arrangement for a doping rate of $x=\nicefrac{1}{3}$ as used in our calculations (see supplementary informations Fig.\,2 left). Our theoretically assumed structures (see supplementary information) would also be in agreement with these diffraction peaks to persist up to relatively high doping levels. At very high doping the doping induced superstructure would then lead to diffraction peaks at one-third and two-third of the original positions in agreement with Fig.\,\ref{fig:struct_doped_30020} (lower right panel). A quantitative analysis of our diffraction data is not possible because of multiple scattering effects.}

\revision{Since the momentum is transfered within the sample plane~\cite{Fink1989}, it is not possible in the present case to directly access the $c$-direction of the crystal.} With the possibility to rotate the sample around an axis perpendicular to the beam direction, measurements are possible in a range of \SIrange[range-units=repeat]{55}{90}{\degree} with respect to this axis. Figure\,\ref{fig:polar_angle} shows elastic scattering measurements (fully doped sample) in a plane rotated relative to the sample plane by the denoted angle. While the left panel shows a clear decrease of the [110]-peak intensity (around \SI{3.6}{\per\angstrom}) with increasing angles, the right panel shows rising features at higher momentum transfers, corresponding to reflections with a finite $l$ component, such as \SI{3.9}{\per\angstrom} representing [114], \SI{4}{\per\angstrom}---[115] and \SI{4.2}{\per\angstrom}---[116]. The corresponding $l$ component of around \SI{0.37}{\per\angstrom} is in agreement with the value of an earlier X-ray investigation of doped \2~\cite{Omloo1970}, where this value was found for a maximum doping rate of $x = \nicefrac{2}{3}$, which is in very good agreement to our stoichiometry analysis (see below).

In order to quantitatively determine substantial structural changes, the behavior of the lattice parameters in the LDA calculation can be summarized by noting that the planar lattice constant stays rather constant upon doping, while the $c$-axis expands with K-intercalation. Figure \ref{fig:Theoretical-results:-a:}b (black solid symbols) shows that in the supercell calculations there is a rather sudden widening of the lattice even on small amount of added potasium, due to the rigid ion cores, which need some space and due to the rigidity of the covalently bound TaSe$_{2}$ prisms. For the cells, where only every second interlayer has K the $c$-axis is smaller (light(red) filled symbols), which follows from the same reasoning. With increasing K-content the $c$-axis shrinks because of the increasing electrostatic interaction provided by an increasing electron content in the interlayer. The behavior of VCA and SC coincide for larger concentrations. However, towards smaller $x$ the $c$ lattice parameter of VCA (open black and light(red) symbols) rises above the SC result, which can be understood from the fact that VCA contains the K$_{18}$-ion cores at a 100$\%$ filling rate for all concentrations, which keeps the TaSe$_{2}$ layers appart, while for supercells all K-atoms vanish for $x\rightarrow0$. This larger $c$-axis spacing however seems to mostly influence the plasma frequency along the $z$-direction, which is not considered in this investigation. However, the $c$-axis value of around \SI{16}{\angstrom} for $x\approx 0.8$ and with it the corresponding $l$ value (\SI{0.39}{\per\angstrom}) are slightly higher than the ones measured and from the literature~\cite{Omloo1970}, respectively.

The planar bare plasmon frequencies of VCA and SC calculations \figref{fig:Theoretical-results:-a:}d form a consistent trend to lower energy
values for increasing potassium content. We emphasize that for undoped \2 the agreement between our measurements and the calculations is
very good. Interestingly, the plasmon frequencies jump, when going from undoped to slightly doped samples, which we attribute to the immediate
lattice widening upon small amounts of intercalated potassium.

In order to gain some insight into the screening due to interband transitions and with it the link to the measured plasma frequencies we also
calculated the real part of the dielectric function in dependence on $x$. In general the real part tends to flatten out into a quasi constant
($\epsilon_{\infty}$) for $\omega<1$eV. Unfortunately, there are some oscillations for small energies due to a low energy feature in
$\mathrm{Im}\epsilon$, whose strength is concentration dependent. This makes a clear extraction of $\epsilon_{\infty}$ rather approximate.
Fig.\,\ref{fig:Theoretical-results:-a:}a shows the trend of $\epsilon_{\infty}$ as a function of $x$. The screening is of the same order as
extracted from experiment and also exhibits the sudden jump, upon the doping-onset. Since $\epsilon_{\infty}$ is not a constant and since the
real part has low energy features we used the theoretically calculated loss function to extract a screened plasmon peak position (Fig.\,\ref{fig:Theoretical-results:-a:}c) and to estimate the K-content of the samples via the measured plasmons peak position. The values
obtained in this manner are used throughout this contribution.

\section{Summary}
\label{sec:summary}

To summarize, we performed \eels\ as well as LDA calculations on the \tmdc\ \2 and thereby investigated the influence of \revision{in situ} intercalation of potassium into the sample. \revision{We demonstrate that our intercalation procedure can be used to achieve rather high doping levels, which allows to study the physical properties of these structures. Furthermore, we present a detailed theoretical analysis of the structure and electronic properties of intercalated compounds.} Upon doping the charge carrier plasmon continuously shifts to lower energies, a behavior supported by the theoretical investigations and ruling out a possible formation of doped phases. \revision{The bands at the Fermi level, which get filled by potassium doping are of Ta\,$5d$ character. Their bandwidth $W$ is rather small (below \SI{2}{eV}), which could hint towards the presence of sizable correlation effects. However, the Wannier functions for these band states as calculated in Ref.\,\onlinecite{Barnett2006} are rather extended, which results in a comparatively small value for the Hubbard $U$. Hence, we expect a (possibly much) larger ratio $\nicefrac{W}{U}$ compared to $3d$ transistion metal compounds and no pronounced Mott-Hubbard physics.}

Structurally, our thin film samples show a transition to a charge-density wave state below \SI{120}{K} in the undoped case and a suppression of the according superstructure upon potassium intercalation. \revision{Moreover}, new superstructures with a \revision{different} periodicity arise that can be assigned to the increasing potassium content in the samples. The changes in the basic structure with respect to the $c$-axis can be consistently found in the \eels\ measurements as well as in the calculations. The spectral width of the charge carrier plasmon significantly decreases upon potassium addition, i.\,e., the plasmon excitation is much less scattered at higher doping levels. Furthermore, as our data indicate the suppression of the charge density wave ground state upon electron doping, it is tempting to speculate about the possibility of \revision{an increasing superconducting transition temperature in potassium doped \2 (\SI{130}{mK} in the undoped compound~\cite{vanMaaren1967})} following the observation that often there is competition between a superconducting and a charge density wave state in transition metal dichalcogenides~\cite{Fang2005,Morosan2006,Hu2007}.

\section{Acknowledgement}

We thank R.\,Schönfelder, S.\,Leger and R.\,Hübel for technical assistance. This work was supported by the Deutsche Forschungsgemeinschaft; grant number
KN393/12 and KN393/13.

\newpage
\section*{Supplementary Material}
\setcounter{figure}{0}
\setcounter{page}{1}
In order to check and complement the virtual crystal approximation (VCA) calculations we constructed a few supercells. No attempt was made to actually determine precise minimum energy structure for each concentration, which can get rather involved due to possible disorder effects or large wavelength structural modulations. We picked cells, which seemed reasonable and have a threefold axis (except for $x=\nicefrac{1}{2}$, see below). For the smallest cell (\SI{50}{\percent} K) we tried three structures in order to get a feeling for the spread of the plasmon parameters $\varepsilon_\infty$ and $\omega_{p_{bare}}$ due to structural and Fermi surface reconstruction effects. For the same reason we additionally considered cells with every second interlayer being empty for $x=\nicefrac{1}{3}$. For $x=\nicefrac{1}{8}$ we only have an alternatingly empty supercell, due to cell size considerations. It turns out that the energies of the alternatingly empty cells are higher but only by a rather small amount of about \SIrange{100}{200}{eV} per TaSe$_2$ formula unit. Such structures could be realized and stable in nature due to kinetic effects during intercalation and the large barrier for K to move through a TaSe$_2$ layer.

\begin{figure}[ht]
  \centering
  \includegraphics[width=0.9\textwidth]{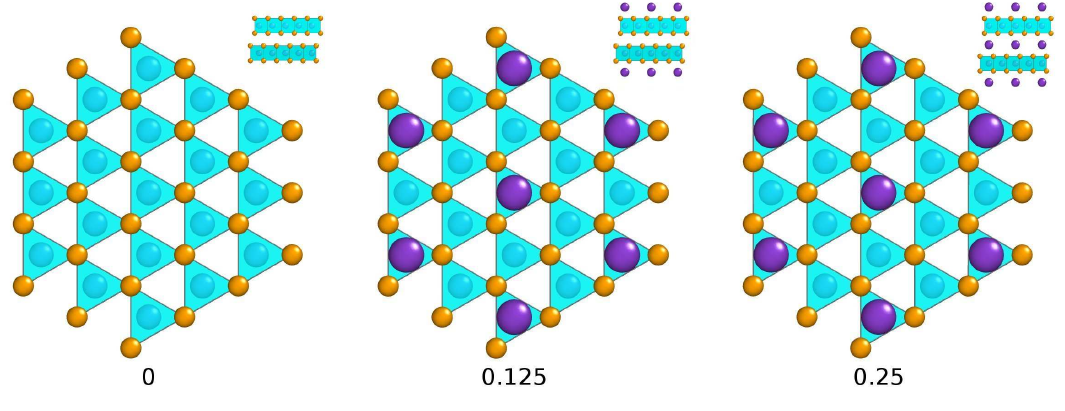}
  \caption{Top and side view of the supercells for $x=0, \nicefrac{1}{8}, \nicefrac{1}{4}$. Atoms at the corner of the prisms are Se, the atoms inside the prisms are Ta and the largest atoms on top are K. The difference between$x=\nicefrac{1}{8}$ and $x=\nicefrac{1}{4}$ is the alternation of empty and filled K-layers for $x=\nicefrac{1}{8}$.}
  \label{fig:1supp}
\end{figure}

\begin{figure}[ht]
  \centering
  \includegraphics[width=0.9\textwidth]{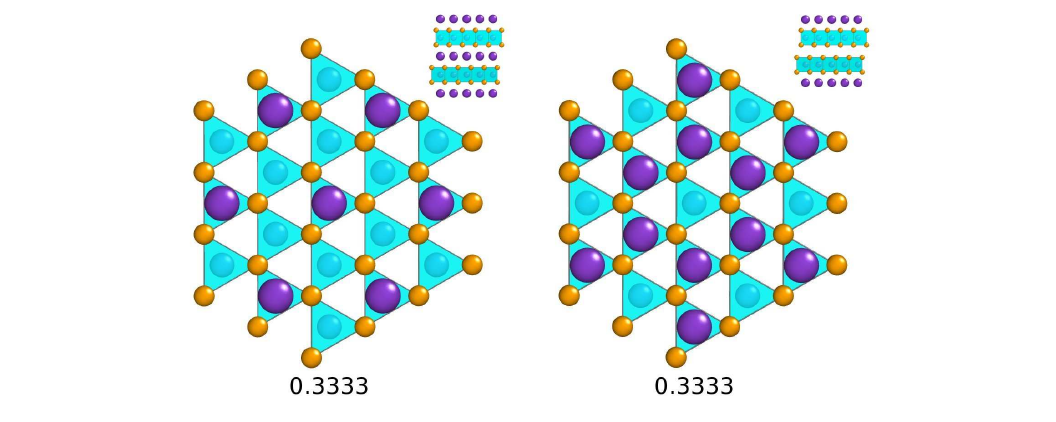}
  \caption{Supercells for $x = \nicefrac{1}{3}$. The cell with every second K-layer being empty is higher in enegry by \SI{73}{mev} per TaSe$_2$ formula unit.}
  \label{fig:2supp}
\end{figure}

\begin{figure}[ht]
  \centering
  \includegraphics[width=0.9\textwidth]{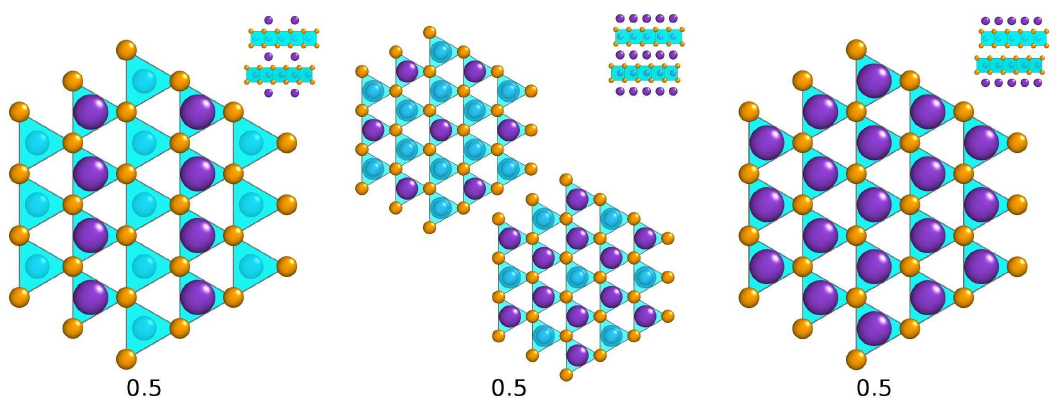}
  \caption{Supercells for $x=\nicefrac{1}{2}$. Note, that the second supercell has two different alternating K-layers. The first cell has orthorhombic symmetry. The first two cellsare energetically degenerate with a difference below \SI{3}{meV}, while the energy of the alternatingly empty third cell os higher by \SI{193}{meV} per TaSe$_2$ formula unit.}
  \label{fig:3supp}
\end{figure}

\begin{figure}[ht]
  \centering
  \includegraphics[width=0.9\textwidth]{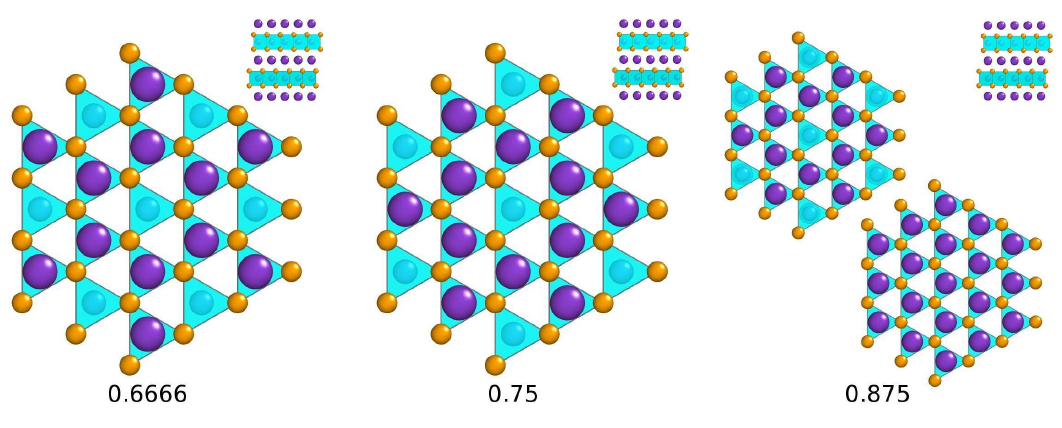}
  \caption{Supercells for $x=\nicefrac{2}{3}, \nicefrac{3}{4}, \nicefrac{7}{8}$. Note, that for $x = \nicefrac{7}{8}$ there are two different alternating K-layers. For $x=1$ (not shown) each TaSe$_2$ prism is covered with one K.}
  \label{fig:4supp}
\end{figure}

\end{document}